\documentclass[12pt]{article}
\usepackage[top=0.85in,left=2.00in,footskip=0.75in,marginparwidth=2in]{geometry}

\usepackage[utf8]{inputenc}

\usepackage{cite}

\usepackage{nameref,hyperref}

\usepackage{microtype}
\DisableLigatures[f]{encoding = *, family = * }

\raggedright
\usepackage{changepage}

\usepackage[aboveskip=1pt,labelfont=bf,labelsep=period,singlelinecheck=off]{caption}

\makeatletter
\renewcommand{\@biblabel}[1]{\quad#1.}
\makeatother

\usepackage{lastpage,fancyhdr,graphicx}
\usepackage{epstopdf}
\usepackage{color}

\definecolor{Gray}{gray}{.25}

\usepackage{graphicx}

\usepackage{sidecap}

\usepackage{wrapfig}
\usepackage[pscoord]{eso-pic}
\usepackage[fulladjust]{marginnote}
\reversemarginpar

\usepackage{amsmath}
\usepackage[utf8]{inputenc} %unicode support

\usepackage{ifpdf}
\usepackage{cite}
\usepackage{array}
\usepackage{mdwmath}
\usepackage{mdwtab}
\usepackage{eqparbox}
\usepackage[caption=false,font=footnotesize]{subfig}
\usepackage{stfloats}
\usepackage[utf8]{inputenc}
\usepackage{array}
\usepackage{makecell}

\usepackage{mathptmx}       % selects Times Roman as basic font
\usepackage{helvet}         % selects Helvetica as sans-serif font
\usepackage{courier}        % selects Courier as typewriter font
\usepackage{type1cm}        % activate if the above 3 fonts are
\usepackage{multicol}        % used for the two-column index
\usepackage[bottom]{footmisc}% places footnotes at page bottom
\usepackage{latexsym}
\begin{document}
\vspace*{0.35in}

% title goes here:
\begin{flushleft}
{\Large
\textbf\newline{Controllability of Social Networks and the Strategic Use of Random Information}
}
\newline
% authors go here:
\\
Marco Cremonini\textsuperscript{1},
Francesca Casamassima\textsuperscript{2},
%Author 3\textsuperscript{1},
%Author 4\textsuperscript{1},
%Author 5\textsuperscript{2},
%Author 6\textsuperscript{2},
%Author 7\textsuperscript{1,*}
\\
\bigskip
\bf{1} University of Milan, Italy
\\
\bf{2} Sogetel, Italy
\\
\bigskip
* marco.cremonini@unimi.it

\end{flushleft}

\section*{Abstract}
This work is aimed at studying realistic social control strategies for social networks based on the introduction of random information into the state of selected driver agents. Deliberately exposing selected agents to random information is a technique already experimented in recommender systems or search engines, and represents one of the few options for influencing the behavior of a social context that could be accepted as ethical, could be fully disclosed to members, and does not involve the use of force or of deception. Our research is based on a model of knowledge diffusion applied to a time-varying adaptive network, and considers two well-known strategies for influencing social contexts. One is the selection of few influencers for manipulating their actions in order to drive the whole network to a certain behavior; the other, instead, drives the network behavior acting on the state of a large subset of ordinary, scarcely influencing users. The two approaches have been studied in terms of network and diffusion effects. The network effect is analyzed through the changes induced on network average degree and clustering coefficient, while the diffusion effect is based on two ad-hoc metrics defined to measure the degree of knowledge diffusion and skill level, as well as the polarization of agent interests. The results, obtained through simulations on synthetic networks, show a rich dynamics and strong effects on the communication structure and on the distribution of knowledge and skills, supporting our hypothesis that the strategic use of random information could represent a realistic approach to social network controllability, and that with both strategies, in principle, the control effect could be remarkable.

% now start line numbers
%\linenumbers

% the * after section prevents numbering
\section*{Introduction}
\label{intro}
{\em Structural controllability} of networked systems~\cite{liu2015control} has been extensively studied since the end of the past decade, following the growing interest in network science~\cite{newman2010networks}. The property of structural controllability is central in the study of how the dynamic of a complex system can be controlled~\cite{lin1974structural}. In short, from control theory, a dynamic system is said to exhibit structural controllability if, with a suitable selection of inputs, it can be driven from one state to any other state in finite time. Inputs to the system are represented by {\em driver nodes} receiving external perturbations. Liu et al. in a seminal paper demonstrated how the problem of determining the minimum set of driver nodes required for structural controllability can be mapped into a maximum matching problem~\cite{liu2011controllability}. Some remarkable theoretical results have been recently demonstrated for complex networks~\cite{menichetti2014network, cowan2012nodal,chen2015paradox,yao2017structural}. 

In complex physical, biological and social networks the same goal of controlling the behavior of large populations of interacting entities has been studied with models of {\em pinning control} regulating networks of coupled dynamical systems. In this case, nodes are dynamical systems coupled according to the network topology. Direct actions are implemented with a certain strength over a subset of selected {\em pinned nodes}, whose reactions propagate into the network~\cite{grigoriev1997pinning,wang2002pinning}. The notion of {\em pinning controllability}, conceptually similar to the structural controllability already mentioned, has been introduced by specifying a reference temporal evolution of the network and two parameters: a coupling gain and a vector of control gains~\cite{sorrentino2007controllability}.  

{\em Adaptive} (or {\em coevolving}) networks are models of complex networks characterized by a mutual interaction between a time-varying network topology and nodes own dynamics~\cite{gross2008adaptive}. These are networks whose behavior is mostly driven by the coevolution of two feedback loops: Nodes dynamics modify the network topology by creating or removing edges (or changing their weight), and at the same time, the modified topology influences the dynamics of nodes. Many real-world complex networks exhibit the coevolution between topology and nodes dynamics, such as in epidemics~\cite{gross2006epidemic,marceau2010adaptive}, opinion and community formation~\cite{iniguez2009opinion}, or socioeconomics networks~\cite{ehrhardt2006phenomenological} and relations among investors~\cite{delellis2017evolving}.

Adaptive networks represent the conceptual framework for this work, which is based on a network model of knowledge diffusion that presents a coevolution between nodes and network topology~\cite{cc-opres-2015}. Nodes interact based on some local preferences and on a question-answer protocol driven by the difference of some key attribute values between neighbors. Interaction between nodes modifies the network topology, which assumes non trivial properties. In its turn, the dynamically evolving topology changes neighborhood relations among nodes.  
It is in this context that we have studied the practical limitations affecting the selection of the theoretically best suited driver nodes - as we will discuss, under certain circumstances it is possible that the only realistic strategy is the opposite of what structural/pinning controllability theory would suggest - and on the type of actions that could be realistically performed in a social context for controlling an adaptive network. 

A relevant theoretical contribution to the problem of controlling adaptive networks, which takes into account physical and economic constraints, has been recently presented~\cite{iudice2015structural}. {\em Structural permeability} is the metric of the propensity of a network to be controlled. Differently from previous approaches and because of this closer to our work, the optimal driver nodes selection is studied in presence of a richer set of constraints: Only some nodes are admissible as driver nodes and some other nodes should not be directly perturbed by control actions. This theoretical framework permits to generalize different problems, including ours (called {\em problem 2} in~\cite{iudice2015structural}), which can be formalized as the selection of the set of driver nodes of minimum cardinality among the set of admissible nodes, satisfying the constraints that controllable nodes are all nodes and no one is excluded from perturbation.  

However, demonstrating structural/pinning controllability for a social network is neither always necessary nor sufficient when realistic scenarios are considered~\cite{motter2015networkcontrology, gao2014target}. It is not strictly necessary because often it is not required to be able to drive a social context from an arbitrary initial state to a specific final state. In most practical situations, there is the need to tune the dynamical evolution from one trajectory driving the population towards a negative outcome to another trajectory, possibly unknown but leading to a better outcome with respect to the welfare. So, in many practical situations, it is not an optimization problem the one we should solve when social networks are considered (i.e., applying the best inputs to the smallest set of driver nodes in order to reach the optimal final state), rather it is a problem of finding heuristics for maximizing the welfare. A possibility is to perturb the system evolution for modifying the {\em basin of attraction} (i.e., modify the dynamics so that the system that was attracted towards a certain space of final states becomes attracted towards a different one)~\cite{cornelius2013realistic,pennacchioli2013three}.
The property of structural controllability is also not sufficient as a requirement, because often there are practical limitations to the type of perturbations that could be injected into the system through driver nodes and also limitations to the accessibility of driver nodes. In many real cases, we are neither free to choose the best inputs nor to observe and manipulate all agents. Most important of all constraints, the individuals subject of the social control observe, judge, and possibly react to the control measures if they are perceived as unethical, unfair, abusive, or oppressive. The long experience with advertising campaigns that ultimately produced adverse reactions (i.e. reactance) \cite{tucker2014social} or the many criticisms concerning the lack of ethics in the social experiment run by Facebook~\cite{kleinsman2015facebook} are well known. The human individual and social rich behavior represents the fundamental difference between control theory for generic complex systems and a control theory for social networks~\cite{vardaman2012interpreting}.
\section*{Goal and Motivations}
Different from structural permeability, in our work we consider two peculiar constraints derived from empirical observations that fundamentally change the approach. The first is on the {\em nature} of admissible direct actions operated on driver nodes, while the second is on the {\em possibility to engage} optimal driver nodes. With respect to the nature of direct actions, we assume that in a social context no explicit direct action controlling the behavior of some individuals is possible unless a contract is enforced, like establishing a sponsorship or signing for a professional service. Indirect actions are instead admissible by exposing individuals to new information, either random information or recommendations, which may or may not produce a change in future activities. We have considered the effects of random information enlarging the knowledge base of driver nodes as a control strategy. With the second constraint, we question the assumption, common in structural/pinning control and network permeability, that the minimal and optimal set of driver nodes could be practically engaged for network control actions. We observe instead that there are situations where this is not the case, one of the more common is when driver nodes charge a fee for propagating control actions to their neighbors, for example if they are celebrities or so called 'influencers' in a social network because of their centrality. In these cases, the price for enrolling a driver node could be  high and budget constraints could restrict their engagement to a sub-optimal number. It is even conceivable that the price tag that network celebrities would impose could be so high that in practice none of the driver node that control theory would prescribe can be actually enrolled. On the contrary, the opposite strategy could become feasible: To enroll as driver nodes a large number of nodes with limited ability to propagate control actions, such as common users and lay people, which however may have unit cost almost irrelevant and could be addressable in bulk through mass campaigns. This is the common scenario for propaganda and mass manipulation through a popular media and, anecdotally, a strategy that has proved effective in several historical occasions. 

The considerations regarding the limitations of structural controllability are the core motivations for the present paper
% which is aimed at studying a realistic social control strategy for social networks based on the strategic introduction of random information into the state of selected driver agents. 
that extends and improves a previous one, in which we had set the basis for the analysis and provided an initial discussion of our hypothesis for a realistically acceptable control of social networks~\cite{casamassima2016use}. Here we focus more explicitly on the two well known strategies for influencing social contexts: The one based on few 'influencers' and the other based on many ordinary users. 

%These approaches have been experimented, respectively, the first in advertising, marketing, and the study of viral phenomena, and the second in recommender systems and search engine studies. Attempts to sustain the circulation of information and to increase serendipitous encounters even through digital interfaces (e.g. browsers) have typically received good acceptance and were perceived as beneficial for the social welfare~\cite{newman2002designing}.
The approaches have been studied in terms of network and diffusion effects, two concepts that we have adapted to our scenario to discuss the rich dynamics produced by the use of random information for social network control. The network effect is analyzed through the changes induced on the network average degree and the clustering coefficient, while the diffusion effect is based on two ad-hoc metrics defined to measure the degree of knowledge diffusion and skill level, and of the polarization of agent interests~\cite{sun2015controllability}.

More specifically, randomness in agent behavior has been modeled as new topics exogenously inserted in agents' state during a simulation: This event wish to represent the typical "unsought encounter" of serendipity and modifies both an agent's criterion of choice of the peers in knowledge diffusion communication. A first study regarding how to adapt the serendipity concept to a social network model has been presented in~\cite{cremonini2016introducing}.
%Results of are produced through simulations of different strategies for either the selection of few hub nodes (influencer agents) or many peripheral nodes (ordinary users) and compared with a benchmark defined by the network behavior without any perturbation. Simulations were also replicated for different network sizes (from small to medium) to verify the effects of scale. 

Finally, the ultimate goal of our work is to demonstrate that exposing selected agents to random information could be practically used to modify the evolution of a social network, nudging it towards behaviors that improve the welfare, and that different strategies to achieve that goal could be realistically elaborated. Examples of problematic behaviors spontaneously emerging from social networks abound, for which the ability to exert a control on the dynamics could be justified. It is well known and documented that in several situations, agents exhibit a strong {\em polarization}~\cite{centola2007homophily, crandall2008feedback}. Agents often form tight communities with few, if any, weak ties between them, heterogeneity of traits and characteristics of connected agents tend to disappear with the increase of homophily~\cite{mcpherson2001birds, golub2012homophily}. In practice, it was often observed how people on social networks tend to slide into "filter bubbles"~\cite{pariser2011filter, easley2012networks} - i.e., self-reinforcing social contexts dominated by information homogeneity with few occasions to have contacts with critical analyses, information from unaligned sources or contrasting opinions - or, in knowledge diffusion and networked learning, how knowledge sometimes diffuses unevenly, for instance agents showing strong tendency towards specialization like when students/recipients of information grow strong interests only in a narrow set of topics disregarding the richness of the full information spectrum~\cite{ACC-compleNet-2014,maglajlic2012engineering}.  
\section{Model}
\label{sec:1}

Our model of social network is inspired by question-answer networks~\cite{zhang2007expertise} where knowledge is shared from expert agents (with respect to a certain topic) answering questions received from less skilled ones. We assume a set of agents and a set of {\em topics} to be given. Each agent has a certain level of {\em interest} and {\em skill} on each topic, both parameters could change interacting with other agents.

More formally, let $N$ be the set of nodes in the network, $T$ be a set of topics and $K$ be a set of time steps. Let 1 and $|K|$ indicate the first and last time step. For each $i \in N$, $t \in T$ and $k \in K$, let $l^k_{it} \geq 0$ and $s^k_{it} \geq 0$ be the level of interest and skill, respectively, on topic $t$ for node $i$ at time $k$. For each $i, j \in N$ and $k \in K$, let $x^k_{ij} \geq 0$ be the amount of interaction between nodes $i$ and $j$ at time $k$, expressed in terms of number of question-answer interactions.

Each agent $i$ has a state characterized by two state-vectors called {\em Personal state $PS^k_{i}$} (what ${i}$ knows at time step $k$) and a {\em Friend state $FS^k_{i}$} (who ${i}$ knows at time step $k$). For each $i \in N$, $t \in T$ and $k \in K$, the \emph{Personal state} has the form $PS^k_{i} $=$ (\bigcup_{j \in T^k_i}(t_j, l^k_{it}, s^k_{it}))$, where $T^k_i \subseteq T$ is the subset of topics that agent $i$ knows at time step $k$. In other words, the {\em Personal state} holds the list of topics an agent knows together with the corresponding level of interest and skill.
The \emph{Friend state} has the form $FS^k_{i}$=$(\bigcup_{j \in N^k_i}({x^k_{ij}}))$, where $N^k_i \subseteq N$ is the subset of agents with whom agent $i$ had interacted up to time step $k$. For each peer $j$ of agent $i$, the number of question-answer interactions $x^k_{ij}$ up to time step $k$ is recorded. 

Together, Friend states define the time-varying adjacency matrix $\mathcal{A}(k)$ of the directed and weighted graph $\mathcal{G}(k) = \{\mathcal{V}, \mathcal{E}^k\}$ describing the network at time step $k$, where $\mathcal{V} := N$ is the set of nodes, and $\mathcal{E}^k  = \bigcup_{i \in N}FS^k_{i}$ is the set of weighted edges recorded by nodes' Friend states.
The element of the adjacency matrix $\mathcal{A}(k)$ is computed as follows:

\[
    a_{ij}(k)=\left\{
                \begin{array}{ll}
                  x^k_{ij} & $if (i,j)$ \in \mathcal{E}^k \\
                  0 & $otherwise$
                \end{array}
              \right.
  \]

\subsection{Network Construction} \label{sub:peerselection}

A network is dynamically created according to the following steps:
\begin{enumerate} 

\item {\em Setup}: At setup (time step 0), for each agent $i$, the {\em Personal state $PS^0_{i}$} has a number of topics $T^0_i$ and corresponding qualities $s^0_{it}$ selected randomly, while the level of interest $l^0_{it}$ is the same for all topics $T^0_i$. The {\em Friend state $FS^0_{i}$}, instead, is empty, because we don't assume any preset network structure.

\item {\em Topic Selection}: At each time step, an agent $i'$ is randomly selected and for that agent a certain topic ($t^* \in T^k_{i'}$) is selected from its Personal state. The choice of the topic is a weighted random selection with values of the associated interests $l^k_{i't^*}$ as weights, this way topics with higher interest are more likely to be selected;
 
\item {\em Peer Selection}: Among agent $i'$ friends and friends-of-friends holding topic $t^*$, the agent $i''$ with the higher skill in topic $t^*$ is selected ($max (s^k_{jt^*} \forall j \in FS^k_{i'} \wedge FS^k_{FS^k_{i'}})$);

\item {\em Successful Communication}: The communication between requester agent $i'$ and the selected peer $i''$ succeeds if the peer is more skilled than the requester in topic $t^*$. The condition for establishing a communication involves the skill parameter: $s^k_{i''t^*} > s^k_{i't^*}$;

\item {\em Failed Communication, Random Selected Peer}: Otherwise, if either Step 3 or 4 fail (i.e., no peer holds topic $t^*$ or the selected peer is less skilled than the requester) then select an agent $i'''$ at random;

\item {\em Communication with Random Selected Peer}: If the randomly selected node holds the selected topic and has a skill greater than that of the requester agent, the communication succeeds ({\em WHILE} $(t^* \in T^k_{i'''}$ {\em AND} $s^k_{i'''t^*} > s^k_{i't^*})$ {\em Communication Succeed}).

\end{enumerate} 

For sake of clarity, it is worth noting that the inclusion of friend-of-friend agents in Step 2 ({\em Peer Selection}) is key for network transitivity and the closure of triangles, two peculiar characteristics of social networks.

The network formation mechanism at {\em start up}, instead, with agents having no friends, is purely random. In particular, considering the six steps just described, a communication attempt always fails at start up because there is no peer to select; Step 5 ({\em Failed Communication, Random Selection}) and Step 6 ({\em Condition for Communication with Random Selection}) are then executed and the initial ties are created by those interactions with randomly selected peers that succeed.

\subsection{Skill and Interest Update} 

After a successful interaction, the agent that started the communication is updated. For simplicity no change in the respondent's state is produced, assuming that knowledge, being an intangible good, does not decrease when shared, and there is no cost of processing and transmission. \emph{Skill} and \emph{interest} are always non negative quantities.
For the skill parameter associated to each topic an agent owns, we assumes that it simply increases in chunks calculated as a fraction of the knowledge difference between two interacting agents. This implies that in subsequent interactions between two agents, the less skilled one accumulates knowledge in chunks of diminishing size. In this simple model of knowledge transmission, skill is a monotonic positive function with  diminishing marginal increments.
For completeness, although not specifically relevant for this work, the model also considers the effect of {\em trust} as an enabling factor for sharing knowledge: The more trust between the two agents, the better the diffusion of knowledge. Here the assumption is that trust between two interacting agents is built with successful communication. In this case, when agents interact for the first time, the chunk of knowledge transferred is reduced by a discounting factor representing the absence of trust. This discount factor progressively vanishes with subsequent communications. This is a simplified form of trust (and distrust) modeling, but motivations could be found in the literature about collective behavior~\cite{Goldstone2009,da2015sudden} and refers both to the prevalence of egocentrism in assimilating new information and to trust dynamics.

The function modeling how agent $i'$ improves the skill associated to a certain topics $t^*$ by interacting with agent $i''$ is: 

\begin{equation} \label{eq:skillexp}
\delta s_{i't^*} =\frac{s_{i'',t^*} - s_{i', t^*}}{\gamma + \rho e^{-\frac{\nu}{\theta}}}
\end{equation}

with:
\begin{itemize} 
\item $s_{i'',t^*} - s_{i', t^*}$ is the difference of skill level associated to topic $t^*$ between agent $i'$ and $i''$;
\item $\gamma \ge 1$ is the control factor for the size of the chunk of knowledge $\delta skill$ that agent $i'$ could learn from agent $i''$;
\item $\rho e^{-\frac{\nu}{\theta}}$ is the discount factor representing trust;
\item $\nu = x^k_{i'i''}$ represents the number of successful communication between agent $n_{i'}$ had with agent $n_{i''}$ at time step $k$; 
\item $-\frac{\nu}{\theta}$ controls the slope of the trust function between the two agents, with $\theta$ the control factor.
\item $\rho$ is the control factor for the actual value of the trust component; 
\end{itemize}

In short, the skill function says that one agent improves its skill with respect of a certain topic by learning from someone more skilled. The improvement is always a fraction of the difference of skills between the two and this passage of knowledge could be influenced by trust relations. From this derives the monotonicity with decreasing marginal gains.
 
The dynamics we have assumed for the \emph{interest} associated to the topic for which the interaction takes place is slightly different with respect to skills. The difference is the assumption of {\em bounded total interest} an agent could have. In other words, one cannot extend indefinitely the number of topics he's interested in or the amount of interest in a certain topic, for the simple reason that time and efforts are finite quantities. Beside the common sense, motivations for this assumption can be found in cognitive science studies, which have shown the tendency of people to shift their attention and interest, rather than behave incrementally~\cite{Goldstone2009}, and in associating the interest for a topic to the time spent dealing with that topic (studying, experimenting, etc.).

With respect to the model and how agent's state parameters are managed, this assumption means that either interest in topics are simply assumed as fixed quantities (a grossly unrealistic assumption) or when the interest associated to a certain topic increases, some other interests associated to all other topics should decrease. We have modeled this second case letting interests to dynamically change, and for simplicity, we assume that the the increase of the interest in one topic is compensated by uniformly decreasing all other interests. A full discussion of these model assumptions was presented in~\cite{ACC-compleNet-2014}.  

 The function modeling the interest function, for agent $i'$ and associated to topic $t^*$ depends on the gain in skill defined by function~(\ref{eq:skillexp}), meaning that the more one learn the more is interested to learn with respect to that topic.
\begin{equation} \label{eq:interestexp}
\delta l_{i',t^*} = \alpha(1 - e^{-\frac{\delta s_{i',t^*}}{\beta}})
\end{equation}

with: $\alpha > 1$ and $\beta > 1$ being the two parameters that control, respectively, the scale and the slope of the interest function, which, again, present diminishing marginal increments. A specific analysis of self-organization strategies based on a manipulation of these two parameters could be found in a previous work~\cite{ACC-compleNet-2014}.

Finally, as mentioned before, to a gain in one interest should correspond a proportional loss in all other interests.

Together these feature of skill and interest functions produce a rich dynamics that exhibits interesting effects with regard to knowledge diffusion into the network. In particular, agents acquire different roles with respect to communication, with a core of tightly connected agents and hubs, and a large periphery. The network tends to form a giant component, but specific setups may broke it into several independent ones. In general, the diffusion of knowledge produced is typically uneven and locally skewed on just few topics. These are the basic characteristics often debated for real case studies with  populations showing a strong tendency to polarization of interests, echo chambers, and forms of cultural isolation even in presence of a social network with a high-degree of connectivity, as the online space now provide.

These are also the key observations that have driven this work, which is focused on studying how, realistically, a social network with the characteristics just presented could be controlled in order to govern the process of knowledge diffusion,  if the goal is to achieve a less skewed distribution of skills, a larger pool of interests shared by agents, and a more effective knowledge diffusion process.

\section{Control Strategies with Random Information}
\subsection{Goal and Motivations}

Based on the model of social network described so far, in this work we study ways to control the dynamic behavior resulting from simulations. Differently from traditional control theory that defines controllability of complex systems as the ability to drive a system from an arbitrary initial state $A$ to an arbitrary final state $B$~\cite{lin1974structural,liu2015control}, for knowledge diffusion in a social network, explicit control mechanisms are typically considered unacceptable, unethical, or severely questioned by the public opinion. This is the case of propaganda, media manipulation, or mass indoctrination. In a social network the notion of controllability must necessarily be weaker and more nuanced than in control theory: It should be an action of nudging the system dynamics with the goal of improving the welfare, rather then an action of tuning a set of inputs that drives a system from any initial state to any desired final state within finite time~\cite{liu2011controllability}.
In this vein, an external controller has a limited set of mechanisms and those mechanisms must be made explicit to the population, otherwise ethical concerns are likely to prevail or the recipients of those influence attempts could revolt against them, as sometimes happened for advertising~\cite{tucker2014social}. With respect to the goal of a control strategy in social networks, it is unlikely to be as precise as in control theory, which is assumed to be able to drive a system from one arbitrary state to any other one. Rather, the realistic goal could be to adjust some macro characteristics of the system, expressed by some {\em metrics}, in order to drive the social network from one {\em basin of possible states} to another one~\cite{cornelius2013realistic,motter2015networkcontrology}. This is, for instance, the way public policies are produced and enforced by public authorities for health, education, employment, or safety reasons.  

This difference is important, because it permits to discriminate control strategies that are not realistic from those that could possibly be considered acceptable in a social context. Acceptability of a control in a social context depends both on the {\em control mechanism} (i.e., how the control is exerted) and the {\em control goal} (i.e., what the control wishes to achieve). In the former case, we exclude from our analysis all mechanisms that involves the use of force upon a population  and all social controls based on deception and trickery. There could be exceptional situations when the use of force is acceptable (e.g., for riots, wars, catastrophes, etc.), but they are out of the scope of this work. In our scenario, a control strategy should always be announced as a public policy to the involved population.
For the latter aspect, the goal of the control in a social context should be aimed at enforcing a public policy to improve the welfare, like reducing inequalities, improving education or social inclusion, or the access to public services. 

More specifically, we assume that a social control strategy is aimed at modifying the dynamics of a social network to drive it from a certain {\em expected basin of resulting states}, if the network is left free to evolve for a certain time and based on present configuration, to a different, {\em desired basin of final states} that increases the welfare. 

In our case study, the control goal is to achieve a {\em better distribution of knowledge among the agents} by reducing polarization, augmenting the average skill level and the average number of topics held by agents.  

The control mechanism, which we consider acceptable in practice, is the {\em strategic use of random information}. Selected agents will be exposed to new random information at certain time steps. The hypothesis is that the controlled exposure to random information, which would not happen if the network is left free to evolve autonomously, can be opportunistically managed to nudge the network evolution. Network metrics and some ad-hoc diffusion metrics are monitored to evaluate the effectiveness of a control strategy.
Examples of a similar idea applied to different contexts can be retrieved in recommender systems~\cite{iaquinta2008introducing,adamopoulos2015unexpectedness} and search engine studies~\cite{bozdag2013bias,white2013beliefs}, which have the related problem of producing recommendations or search results too heavily influenced by the personal profile of the requester.

\subsection{Network and Diffusion Metrics}
To evaluate the effectiveness of control strategies, we consider how exposing a subset of agents to random information may influence some basic metrics. This represents our interpretation of {\em acceptable control mechanism applied to a social network}: Some key parameters of the network topology and its dynamics are influenced in a predictable way and change the original evolution trajectory. The structure of the network might be modified, as well as the relevance of some agents with respect to the connectivity or the diffusion of knowledge; it could also be reduced the polarization of agents and the average number of topics that agents learn from interactions with others. In short, an acceptable control mechanism should be able to influence in a predictable way (at least stochastically) the coevolution of the adaptive network.

In this work, we evaluate two classes of metrics: {\em network metrics} and {\em diffusion metrics}. Among the many network metrics defined in the literature, we focus the analysis on the {\em average degree} and the {\em clustering coefficient}. Together these two metrics provide many information about the network topology and the communication infrastructure. For instance, the type of network can be inferred, as well as the connectivity distribution between the core and periphery~\footnote{Network {\em core} and {\em periphery} are typically defined in terms of assortative mixing. For assortative networks, as social networks typically are~\cite{newman2003social}, high-degree nodes tend to link together forming a group of higher mean degree than the network as a whole. This subnetwork is called the core. Low-degree nodes, instead, being the majority of nodes, form a generic periphery of lower mean degree than the core~\cite{newman2002assortative,newman2003mixing}.}. Some agents typically becomes hubs and exert a strong influence over the communication and distribution of information in the network. The ability to modify network metrics such as the average degree and the clustering coefficient has influence over the hub agents, for instance their prominence could decline in favor of a more decentralized communication.

The second class of metrics are defined ad-hoc. We have called them diffusion metrics because their aim is to measure some important attributes of the spread of topics and interest among the agents.  
Two metrics have been defined: {\em Average Knowledge (AK)} and {\em Knowledge Diffusion (KD)}, used to evaluate different characteristics of the knowledge diffusion dynamics.

% {\em CE} measures how often agents are able to successfully interact with others with respect to the number of requests they made during a simulation.
% %
% \begin{align}
% CE =\frac{\sum_{i=1}^{N}\sum_{j \in N_i}x_{ij}}{|K|} 
% \end{align}

% With $N$ the total number of agents, $|K|$ the number of requests attempted by all agents, one for each time step, and $x_{ij}$ already defined as the number of successful question-answer interactions between agents $i$ and $j$ (for simplicity of notation, we omitted the attribute $k$ related to the time step of the calculation). {\em CE} is a measure of performance and a conversion rate, which provides a general estimate of the efficiency of agents in their communication attempts. $CE = 0$ means that there is no communication in the network since every interaction attempt failed; $CE = 1$ means instead that every interaction attempt succeeded.

{\em AK} is calculated as the {\em average skill with respect to the topics actually owned by agents}. In this sense, $AK$ contributes to measure the specialization/polarization of agents, because $AK$ could be high when agents knows few topics with high skill levels.  

{\em KD}, instead, is the average skill of agents {\em with respect to the total number of topics in the network}. This means that, implicitly, a topic that an agent does not own is averaged with skill zero. The meaning of $KD$ is to evaluate the actual diffusion of knowledge with respect to the case of perfect diffusion, when all agents hold all topics with maximum knowledge. In this sense, $KD$ is relevant because it gives an indication of how effectively topics are spread over the population and with which skill level.

\begin{align}
AK = \frac{\sum_{i=1}^{N}\sum_{j=1}^{|PS_i|} s_{i,j}}{\sum_{i=1}^{N}|PS_i|} 
\end{align}

\begin{align}
KD = \frac{\sum_{i=1}^{N}\sum_{j=1}^{|PS_i|} s_{i,j}}{N \times T} 
\end{align}

With $|PS_i|$ the number of topics of agent $i$'s' Personal state and $T$ the total number of topics in the population. 
\subsection{Social Control Strategies}

For the social network control based on random topics, we assume that, with a given frequency, a certain subset of agents is selected and a variable number of random topics, among the ones existing in the network, is inserted into their Personal state with an average level of interest. The goal is to force the exposure of some agents to a larger set of topics and exploit the mechanisms of topic and peer selection described in Section~\ref{sub:peerselection} to modify the dynamic network behavior. With regard to network structure, with new random topics, the network construction mechanism strictly based on topology is occasionally bypassed and a bridge effect is produced. 

Recalling the terminology typical of control theory, we call {\em driver agents} the agents selected for being exposed to new random topics. The key parameters for defining a social control strategy are: The {\em selection criteria of driver nodes}, the {\em number of random new topics introduced}, and the {\em frequency of introduction of new topics}.    

The simulations and results discussed in the following will be based on different choices of these parameters. We made some  assumptions with respect to the parameters choice. For the selection criteria of driver agents, in this work we focused on just two particularly relevant options: Driver agents are selected as a fraction of total agents {\em ranked based on the node degree, either in decreasing on increasing order}. 

\subsubsection{Selection of high-degree driver agents}
In this case, driver agents are a certain subset of agents with higher degree (1\% or 10\% are the values used in tests). This means that agents exposed to new topics, and whose behavior will be directly perturbed, are those that mostly centralize the communication in the network. They are the hubs and the nodes of the network core, typically densely connected. Therefore, with this choice, very intuitive and similar to strategies adopted in social media and advertising (e.g., targeting the network influencers), it is the core of the network and the hub nodes to be directly manipulated. With respect to our metrics, we expect that few nodes will have a strong influence over the network dynamics and that hub nodes will increase their degree. Overall, though, this should stimulate a better communication, with positive effects on the knowledge diffusion.  

\subsubsection{Selection of low-degree driver agents}
In this other case, instead, driver agents are chosen among those with lower degree. This option may seem counterintuitive, being these agents the least connected nodes of the periphery of the network, those that typically have the least influence on communication and spreading processes. It has, however, some interesting properties. At network level, targeting the periphery of the network means to directly stimulate the closure of more triangles. This means to raise the clustering coefficient and foster a more rich decentralized communication, which may have effects on the network topology. Regarding diffusion, these nodes are likely those that had fewer connections, typically connected with other low-degree nodes (i.e., social networks are typically assortative~\cite{barrat2004architecture}, although some recent experiments have showed that during the evolution a network could transit from assortative to disassortative~\cite{li2014sparse}), thus they have a limited choice of peers for exchanging knowledge. Exposing them to random topics very likely profoundly change their pattern of interaction, much more than what happen to hub nodes, which were and remain highly connected. On the other side, the influence that a typical low-degree node has on communication is very limited, for this reason the number of low-degree agents that must be selected to produce a clear control effect on the network is much higher than in the previous case. In our examples we have showed results with 50\% and 70\% of agents selected as driver agents. The choice of 50 and 70\% is qualitative and motivated by presentation simplicity. The goal here is to show that, in principle, targeting ordinary users instead of influencers may produce relevant effects on the coevolving network dynamics, but whether in one case the driver nodes are a small subset with respect to the whole network (potentially requiring a high price to be engaged), in the other case the number of driver nodes is on a different scale, likely a large subset of nodes (possibly with low unit cost).

% we show that  50\% rate of low-degree driver agents is needed in order to produce effects on metrics comparable with the 1\% of high-degree agents and the 70\% rate of low-degree agents to compare with the 10\% of high-degree agents.

% In practice, this strategy is the opposite of that targeting influencers and famous individuals. It targets average users and ordinary people, each one of these agents has very limited diffusion capacity, but the social control strategy could be very effective on these type of agents and, by adjusting the behavior of a large part of them, the resulting effect on the whole network could be relevant.

\subsubsection{Number of new topics}
The second important parameter for defining a social control strategy represents the amount of perturbation that we are introducing in driver agent's Personal state. Instead of an actual number of topics, it is expressed as a rate of new topics a driver node will receive with respect to the ones already present in its Personal state. Typical values we have considered are 1\%, 10\% or 30\%, defined empirically as simply the rate that, on average, gives just one new topic and two rates that permit to incrementally add new topics during a simulation. For simplicity, we have assumed that all driver agents receive the same rate of new topics. Alternatives would have been possible, like simulating fixed amounts of new topics (we have tried this approach, but preliminary results were not as interesting as for the fixed rate option), or adjusting the number/rate of new topics based on contextual information for each agent (we did not explore this possibility, because it further complicates the model and the dynamics, but it is promising for developing a "personalized control strategy").

\subsubsection{Frequency of new topics}
The considerations for the frequency of introduction of new topics are similar to what discussed for the number of new topics. In this work, for simplicity we have considered fixed time intervals (measured in number of simulation steps, called {\em ticks}). 1000 ticks is the value used in the examples, for simulations that typically span over 100000 ticks. As in the previous case, the value is selected empirically, mostly for sake of presentation. Again, similar to the previous case, more elaborate time-based strategies were possible and interesting to study. The action of inserting new topics, instead of being a fixed frequency, could be triggered by contextual information, like a particular state of the network or particular values of the metrics, for instance when the overall communication stagnates. Another, even more elaborate strategy, could be to trigger the addition of new topics at agent level, when the specific driver agent exhibits a particular state or behavior (again, another form of "personalized control strategy"). 

\section{Model Instances and Simulation Results}

\subsection{Basic Settings}
The model of knowledge diffusion with social control mechanism is implemented in Netlogo~\footnote{https://ccl.northwestern.edu/netlogo/}, a multi-agent programmable modeling environment, which has a proprietary script language and permits to run simulations of complex networks through the GUI or the API. 

All simulations we present here have the basic setting listed in Table~\ref{tab:instances} ({\em Parameters shared by all instances}), chosen mainly for presentation sake, but still representative of network behaviors observed during tests with a large number of settings.  

For the dynamics produced by our model, the rate between the number of agents and the number of topics $\frac{N}{T}$ is key, because it directly influences the average number of topics the agents are assigned at setup and the ability to select peers. It is to simplify the analysis and the presentation that we fixed the number of topics $T$ and the maximum number of topics assigned to agents at setup $\lambda_T$, and varied the number of agents $N$ to analyze, at least partially, how larger populations affect the results (see Table~\ref{tab:instances} ({\em Network sizes})). In this study, we focused on small-to-medium sized populations, possibly representing  situations like network learning for traditional academic courses (or even local subgroups of participants to large online courses interacting in dedicated forums), or online social networks built around restricted cultural or professional contexts. Very large communities represent a structurally different case study, which needs a specific analysis, out of the scope of this study. The duration of simulations, defined as the maximum number of time steps $|K|$, instead, has been decided empirically as the number of time steps sufficient for networks produced by our model to reach a steady state. 

\subsection{Model Instances}
Model instances presented in this work are aimed at discussing how the two strategies for driver agents selection, one selecting a subset based on agents with higher node degree and the other agents with lower node degree, could be exploited for controlling the network behavior. The two options have been compared also with the case of network evolving without any perturbation.
The choice of node degree as the attribute for selecting driver agents has several references in control theory literature and is an attribute typically analyzed in social network studies. Several alternatives to degree centrality exist for selecting driver agents, like betweenness and eigenvector centrality, or more elaborate criteria based on community detection. However, we feel that before exploring alternative strategies and study which one could optimize the outcome, the goal should be to acquire a better understanding with respect to the possibility to use random information as a social control mechanism in practical situations. The simple criteria based on node degree has the important benefit of being intuitive in practice (i.e., selection of leaders/influencers or of common users/participants) and already exploited for different applications. 

Specifically, for the case of high-degree nodes, driver agents are either the 1\% or the 10\% of the population (see Table~\ref{tab:instances} ({\em High-degree agent selection})). For low-degree nodes, driver agents are either the 50\% or the 70\% of the population (see Table~\ref{tab:instances} ({\em Low-degree agent selection})). As previously discussed, the different characteristics of high- and low-degree nodes with respect connectivity and controllability makes the comparison of equally sized subsets not meaningful. 

\bgroup
\def\arraystretch{1.5}
\begin{table}[ht!]
\begin{center}
\caption{\label{tab:instances}Parameters for the model instances.}
\begin{tabular}{ |l| }
\hline
%\makecell[l]
\thead{{\bf Parameters shared by all instances}} \\
\hline
{\em Total Number of topics in the system:} $T= 100$ \\
{\em Max Number of topics per agent at setup:} ${\lambda}_{T}= 10$ \\ 
{\em Max Number of time steps (max \#ticks):} $|K|= 100000$ \\
{\em Number of New Random Topics per driver agent (\% over $T_i$):} 30\% \\
{\em Time interval of New Random Topics Insertion (in \#ticks):} 1000 \\
\hline
\thead{{\bf Network sizes}} \\
\hline
%\makecell[l]
{\em Number of agents:} $N= [100, 200, 500, 1000]$ \\
\hline
\thead{{\bf High-degree agent selection}} \\
\hline
{\em Number of Driver Agents (\% over $N$):} [1\%, 10\%] \\ 
\hline
\thead{{\bf Low-degree agent selection}} \\
\hline
{\em Number of Driver Agents (\% over $N$):} [50\%, 70\%] \\ 
\hline
\end{tabular}
\end{center}
\end{table}
\egroup

\subsection{Network and diffusion effects}

With simulations we wish to show two distinct effects that together contribute to define the effectiveness and feasibility of a social control strategy based on random topics: the {\em network effect} and the {\em diffusion effect}.

Network effect describes how the addition of new random topics permits to modify the topology structure of the social network. This effect is one of the most fundamental for a social control strategy, because by interfering with the process that produces the network topology, some critical parameters of the network structure are affected, like the degree and type of assortativity, the size of independent components, the weight and number of hub nodes in centralizing the overall communication, and the relation between the core and the periphery of the network. From these parameters many emerging behaviors and features depend, for instance the community structure together with the prevalence of segregation or of echo chambers, the robustness with respect to failures, attacks, or deception, and the speed and the spread of a diffusion phenomena~\cite{lu2016attack}.
The network effect is measured by typical social network metrics like centrality metrics, homophily degree, density degree, the analysis of structural holes and of tie strength among the others.
In this work we focused on two metrics among the most basic ones: {\em average node degree} and {\em clustering coefficient}, which let to consider some fundamental aspects of network topology.

With diffusion effect, instead, we mean how the new random topics modify the specific mechanism of knowledge transmission defined by our model. Our knowledge transmission is ad hoc, but it wishes to represent a general social context where agents have different skills and different willingness to acquire new knowledge. Skills are improved based on personal one-to-one contacts supported by the{} network topology. This effect is the one interfering with the topic selection mechanism of driver agents, which in turn affects the peer selection criteria and consequently changes how likely is a driver agent to maximize the skills on few topics only or to distribute the interactions on an extended set of topics. Also, the diffusion effect influences the average communication efficiency of agents, because by easing the creation of new interactions, communication congestions and spikes are less likely. 
We measure the diffusion effect with the two previously defined ad hoc metrics called {\em Average Knowledge (AK)} and {\em Knowledge Diffusion (KD)}. {\em AK} represents the average skill level of agents with respect only to the topics they possess.
{\em KD} measures the average skill level of agents with respect to all possible topics. Then, by design, {\em AK} is an upper bound for {\em KD}.

In particular, {\em KD} necessarily starts very low at the beginning of a simulation because in our settings the number of topics assigned to agents at start up is limited and the total amount of skill is bounded (i.e., in these simulations, on average, agents at start up have 5\% of the total number of topics $T$). Then {\em KD} monotonically increases with the exchanges of knowledge among agents and the diffusion of topics, up to a value $max(KD)$, which depends on the specific settings. It represents a measure of the {\em erudition of the social network} and gives an indication of how effectively skills are spread over the population and the level of knowledge. For our analysis, it is important how fast it approaches {\em AK}.

On the contrary, the meaning of {\em AK} depends on the run time network state and it can either increase or decrease. For the same reason {\em KD} always starts low, {\em AK} always starts with a value close to the middle of the scale (i.e., at start up agents have a fixed total amount of skills spread over few topics). From that point, {\em AK} measures how much the proportion of the two quantities varies. It stay constant if the two change proportionally during the execution; it increases when agents acquire skill level faster than they acquire new topics; and it decreases {\em vice versa}. 

Given these ad hoc metrics, we say that an agent is {\em polarized} when it concentrates its best skills on just very few topics. {\em Polarization of the social network} is another important global characteristics, together with erudition. Taking into account {\em AK} and {\em KD} together, the degree of network polarization can be evaluated. We say that {\em a network tends to polarize} when the difference between {\em AK} and {\em KD} is {\em increasing}, because it means that agents are gaining skills mostly on the same subset of topics without a corresponding increase in erudition. On the contrary, {\em a network tends to heterogeneity} when the difference between {\em AK} and {\em KD} decreases, meaning that agents are starting to spread their interests on more topics.

\subsection{Simulation Results}

Results of simulations are presented in Figure~\ref{fig:network} and Figure~\ref{fig:diffusion}, which show, respectively, the network and the diffusion effects of the strategic use of random topics.

For the network effect, Figure~\ref{fig:network} presents how the average node degree and the clustering coefficient change with respect to the benchmark represented by the typical network behavior with no injection of random topics in nodes (label {\bf C0} in Figure~\ref{fig:network}) for three network sizes ($N=[100,500,1000]$). 

The settings involving the addition of random topics are the ones previously introduced and whose parameters are summarized in Table~\ref{tab:instances}:
\begin{itemize}
  \item Label {\bf HIGH 1\%} represents the selection of the 1\% of high-degree nodes as driver agents; 
  \item Label {\bf HIGH 10\%} represents the selection of the 10\% of high-degree nodes as driver agents; 
  \item Label {\bf LOW 50\%} represents the selection of the 50\% of low-degree nodes as driver agents; 
  \item Label {\bf LOW 70\%} represents the selection of the 70\% of low-degree nodes as driver agents; 
\end{itemize}

As discussed, the two general strategies tested could be interpreted as the selection of influencer agents the one labeled as HIGH and the selection of ordinary users the one labeled as LOW. 
The choice of the rate of driver agents is mostly for presentation sake. However, we note that the values of the selection of high-degree nodes are aligned with the ones for network controllability of social communication networks analyzed in~\cite{liu2011controllability}. 

\subsubsection{Network effect results}
\begin{figure}[ht!]
	\includegraphics[width=\linewidth]{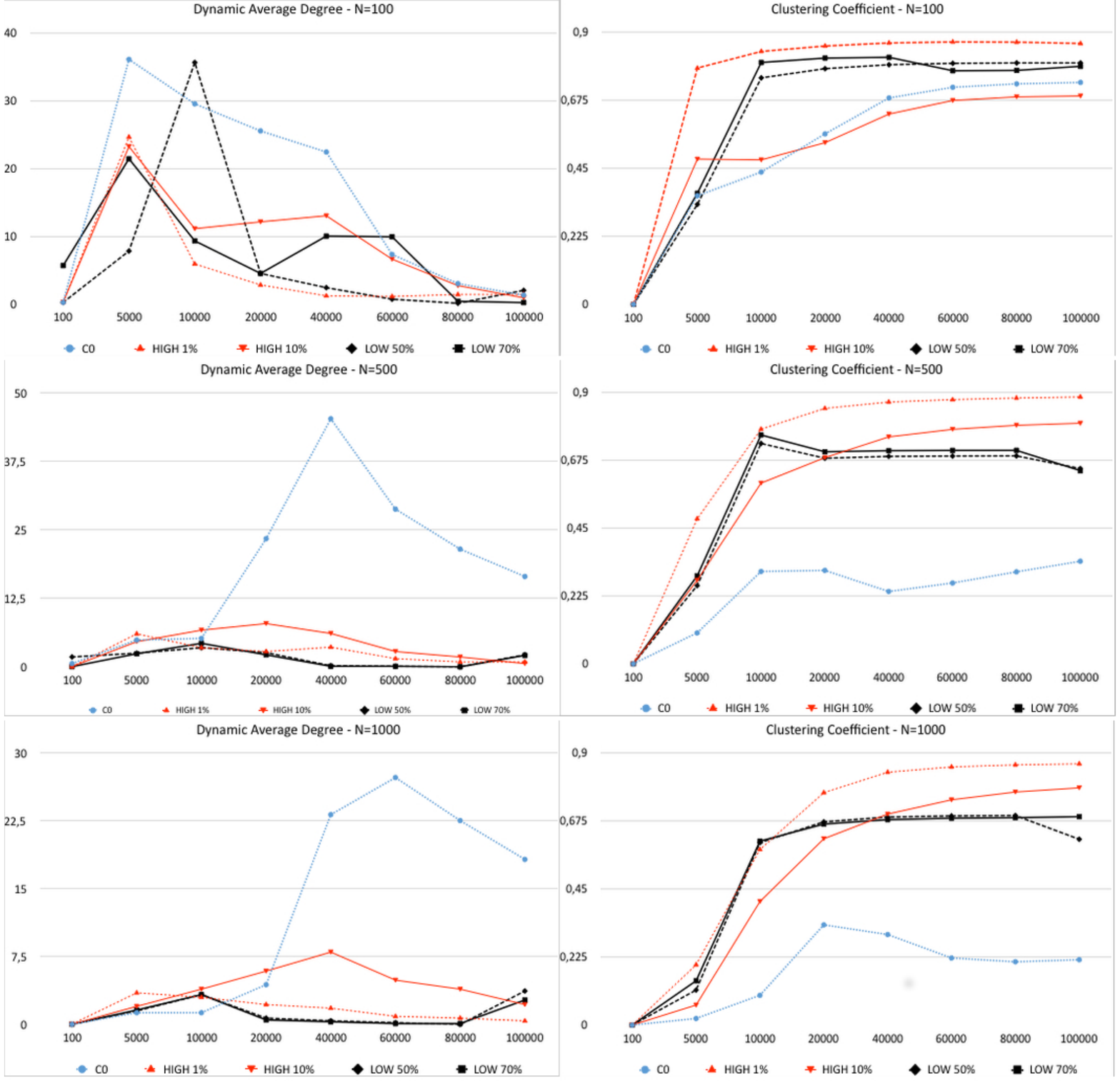}
   \caption{\label{fig:network}Network results for the selection of high or low-degree driver agents.}
       {\em Dynamic Average Degree} and {\em Clustering Coefficient} simulation results for {\em N = [100, 500, 1000]}.
       Label {\em C0} represents results of the natural system dynamics with no addition of random topics. {\em 1\% HIGH} and {\em 10\% HIGH} labels represent results of, respectively, the 1\% and the 10\% of the {\em high-degree nodes}. {\em 50\% LOW} and {\em 70\% LOW} labels represent results of, respectively, the 50\% and the 70\% of the {\em low-degree nodes}. The x-axis represents time steps as ticks of the simulation; y-axis represents the average degree ({\em left}) and the clustering coefficient ({\em right}).  
       
       \end{figure}

For the \emph{average degree}, the results of Figure~\ref{fig:network} present what is called {\em dynamic average degree}, a representation specific for dynamic network behaviors, which calculates the average node degree taken on certain time step periods (e.g., in Figure~\ref{fig:network}, time step periods are [0,100], [100,5000], [5000,10000], [10000,20000], [20000,40000], [40000,60000],[60000,80000], and [80000,100000]). The value for each time step period is the average degree corresponding to active links (new or reused) in that specific period~\footnote{The values of the dynamic average degree are not calculated with the traditional formula $<d>=\frac{2*E}{N}$, with $E$ the number of links and $N$ the number of nodes in the network. Instead, for each time step period $(t_{i-1},t_i)$, with $E_{t_i}$ the number of active links in $(t_{i-1},t_i)$, the dynamic average degree for each time period is $<d_{t_i}>=\frac{2*E_{t_i}}{N}$.}. 

The form of the dynamic average degree function produce a rich representation of network behaviors. One aspect that clearly emerges is the spiky behavior of the original network {\em C0} that tends to quickly saturate all possible connections.
The injection of random topics, instead, in all situations, even for very few driver nodes (e.g., for $N=100$, the setting {\em HIGH 1\%} means that the node with the highest degree is the only driver agent) is able to produce a sensible reduction of the average degree with respect to the benchmark {\em C0}. 

In tests labeled {\em HIGH}, driver agents certainly include hub nodes and the large influence these nodes have on the network structure is clearly demonstrated by the fact that it suffices to modify the state of a few of them to obtain a wide effect on the degree metric. The average degree drops from 50 to 80\%. This is a strong network effect and potentially a powerful way of controlling a network behavior, because it signals a reconfiguration of the communication within the network.  

At first sight, a strong reduction of average node degree might be interpreted as a sensible reduction of the weight of hub nodes. However, inspecting the outcome at agent level, it could be observed that this is not the case. The average degree of hub nodes actually {\em increases}, which is reasonable giving the fact that they receive new topics. The global reduction of average degree is driven by the average degree reduction in the large periphery of the network. The communication is even more centralized by few hub nodes and overall the network tends to become disassortative, similar to the effects discussed in~\cite{li2014sparse}.

With respect to the \emph{clustering coefficient}, the tests show a general increment compared to the benchmark. This signals that more triangles have formed, a direct effect of the new connections produced by the introduction of random topics in hub nodes. Together, the reduction of the average degree and the increase of the clustering coefficient tends to enhance the small-worldiness of the network, a characteristic previously not present, which is an important effect in terms of controllability~\cite{kossinets2006empirical,newman:2003}.

Considering the different network sizes, we observe that the degree reduction is systematic, with results very similar for the larger networks. The case of $N=100$ represents a peculiar situation because in that setting the communication efficiency is very high and agents establish quickly many connections. This is the reason for the initial spikes of the average degree in all situations. 

Another interesting result that emerged from these tests is that with respect to the ability to influence network metrics, both strategies {\em HIGH} and {\em LOW} could be effective. However, to have comparable strength, the number of driver nodes must be necessarily very different in the two cases, as showed in our simulations. 
However, there is an important difference between the {\em LOW} and the {\em HIGH} strategy with respect to hub nodes. By selecting low-degree nodes as driver agents, the formation of hub nodes is more difficult and the communication becomes more decentralized and local. The effect on the communication structure of the {\em LOW} strategy is the opposite of the {\em HIGH} one. This difference appears very clearly in the following diffusion effect.

In general, these observations present an interesting practical control problem still not fully investigated in the literature. While some theoretical results have been studied for reference network topologies~\cite{cornelius2013realistic, menichetti2014network, liu2011controllability}, few studies exist that consider mechanisms for dynamically adjusting the degree of control during the evolution of a social network. 

\subsubsection{Diffusion effect results}
\begin{figure}[ht!]
  \includegraphics[width=\linewidth]{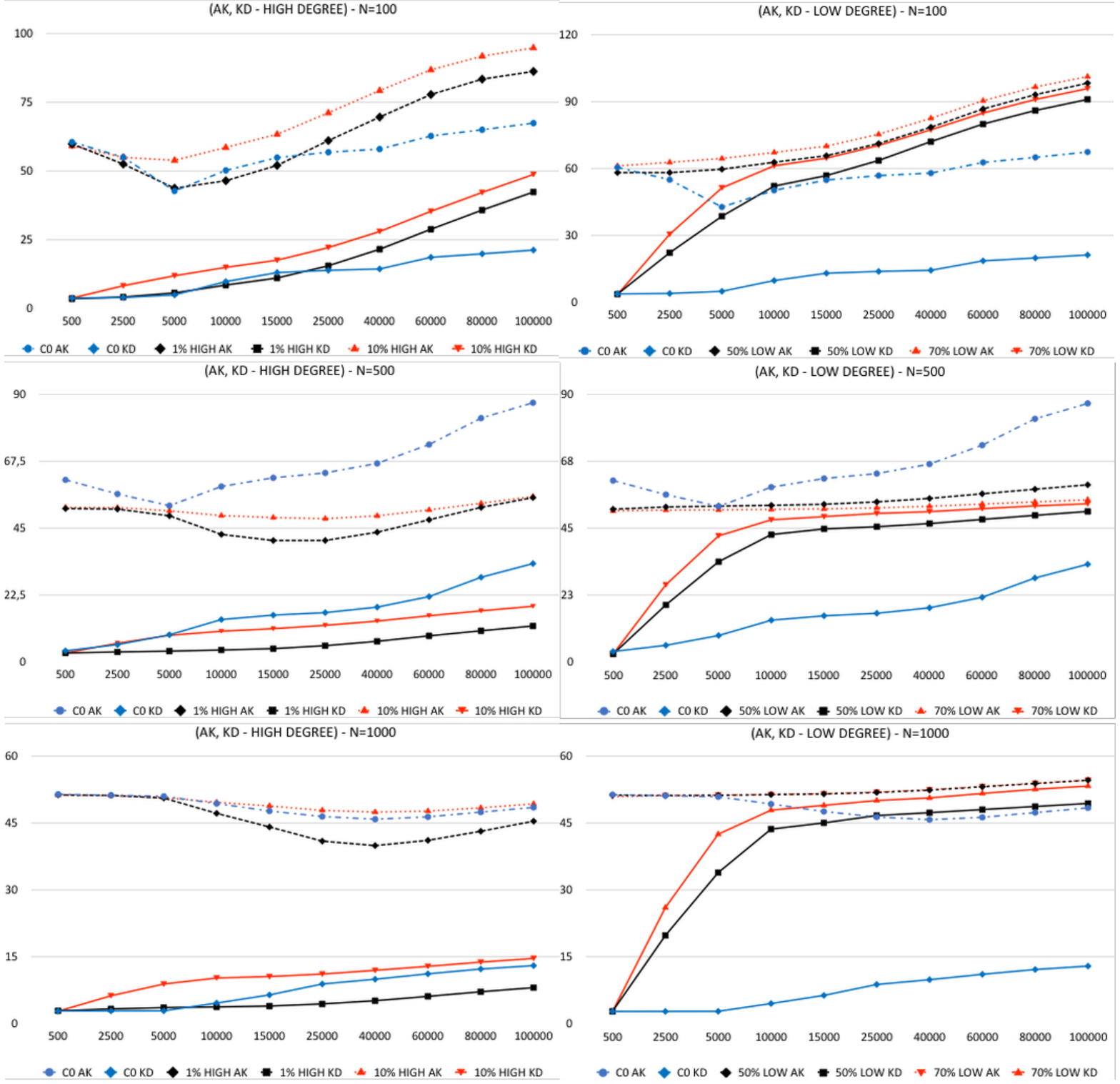}
   \caption{\label{fig:diffusion}Diffusion results for the selection of high or low-degree driver agents.}
       {\em Average Knowledge (AK)} and {\em Knowledge Diffusion (KD)} simulation results for {\em N = [100, 500, 1000]}.
       Label {\em C0} represents results of the natural system dynamics with no addition of random topics. {\em 1\% HIGH} and {\em 10\% HIGH} labels represent results of, respectively, the 1\% and the 10\% of the {\em high-degree nodes}. {\em 50\% LOW} and {\em 70\% LOW} labels represent results of, respectively, the 50\% and the 70\% of the {\em low-degree nodes}. The x-axis represents time steps as ticks of the simulation; y-axis represents the value of {\em AK} and {\em KD}, which always ranges in [0,100] for these tests.
       
       \end{figure}

With these tests we have studied how the control strategies affected the diffusion of topics and of skills in the agent population. Here the goal is to modify the dynamics in order to reduce the polarization and increase the erudition of the network. The results are showed in Figure~\ref{fig:diffusion} and describe two very different scenarios for {\em HIGH} and {\em LOW} strategies.
With respect to the {\em HIGH} strategy, the results do not provide evidence of a clear improvement compared with the benchmark. For $N=100$ there is a slight improvement in the actual values of {\em AK} and {\em KD}, meaning a larger diffusion of topics and skills (higher erudition), but same high level of polarization. For $N=500$ the outcome is opposite: Actual values are worse than the benchmark, thus a reduced diffusion of topics and skills, but also a smaller level of polarization. For $N=1000$ the results of both 1\% and 10\% {\em HIGH} are very close to {\em C0} for both {\em AK} and {\em KD}.
In this regard, selecting the few hub nodes as driver agents do not seems particularly effective for a better diffusion of knowledge measured as increased erudition and reduced polarization.

Opposite is the outcome of the {\em LOW} strategy. In this case, it appears a systematic improvement with respect to the benchmark. For all network sizes and in both settings, 50\% and 70\%, {\em KD} approaches {\em AK} very fast as a result of the addition of new random topics to many nodes. This is a clear sign of no polarization, meaning that agents are able to distribute their skill over the enlarged set of topics. For $N=100$ there is also a clear improvement in absolute values of {\em AK} and {\em KD}, while for $N=500$ and $N=1000$ {\em AK} grows slowly or tends to stay flat. This behavior signals that skills increase slowly and with difficulty. The reason is that many nodes with low-degree receive new random topics, but, after the initial burst of activity with neighbors, most of the time they are unable to find a peer within the few friends or friends-of-friends. Overall, for $N=100$ we are able to both improve erudition and have no polarization, the optimal combination for the welfare. For larger networks, instead, adding new random topics quickly removes polarization because locally the communication flow efficiently, but the positive effect is then limited by the fact that with few hub nodes in the network, agents are unable to easily find new peers.

To summarize, with these tests we have showed two possible effects of a social control strategy based on the addition of new random topics to selected driver agents. The results are not conclusive, but we have presented some important differences between addressing the influencers or the ordinary users. We have also seen how important is to define the goal of a control strategy, because the ability of manipulating some network metrics does not necessarily imply to be able of improving context-dependent features like the diffusion of knowledge, and {vice versa}.

\section{Conclusions}
In this paper we discussed a possible use of random topics as control inputs for driver nodes, based on an adaptive network model for knowledge diffusion, and two realistic control strategies, one based on high-degree nodes commonly analyzed in studies of structural/pinning controllability, the other based on low-degree nodes, which instead had received few attention in controllability studies, but has practical application in several situations. 
The idea of using random information in control strategies has a long tradition in control theory and could be also retrieved in recommender systems and search engine research, which share the same problem of excessive polarization of interest and the need to improve diversity that we consider for knowledge diffusion~\cite{bradley2001improving, adomavicius2012improving}. However, explicitly considering random information as a control input for social networks is a new idea worth exploring, in our opinion. The two general strategies for influencing the coevolving dynamics are based, respectively, on the selection of few influencers and the manipulation of their behavior in order to drive the network; and on the selection of many ordinary users as the driver of the network as a whole. The two approaches have been studied in terms of network and diffusion effects.

The problem of controlling social networks presents striking differences with respect to the study of structural controllability for generic complex networks. The social context introduces many limitations (ethical, operational, functional), and often does not strictly require full structural controllability, rather the ability to nudge the network evolution towards a certain basin of results improving the welfare.
For these reasons, the application of control theory to social networks could be a source of inspiration, but it cannot be applied as-is, because important adaptations are required, which represents a promising and still largely unexplored research strand.  

There are many situations in which it would be important to know how to handle the level of random information that agents possibly receive. For instance, in learning situations, in social media, journalism, knowledge diffusion, skill acquisition, experience dissemination, epidemics, and possibly risk management. In all these situations, there could be the problem of an excessive polarization (of interests, attention, analyses) and lack of erudition, but the solution cannot be to simply change what individuals prefer or believe or regard as important/interesting. Increasing information heterogeneity and serendipity could be effective approaches for improving the controllability of social contexts.  

Furthermore, considering that in practical situations it could be impractical to either recognize all theoretical driver nodes, accessing them with external perturbations, or sustaining the cost to engage them, we have presented some empirical solutions based on network metrics for selecting nodes that might have practical usage. The results of our work look promising to us and encourage more analyses, tests, and verifications with respect to real social networks. Other future works will include a theoretical grounding to the results, the definition of an analytical model and a statistical analysis of a larger test set to better evaluate the performance of control strategies. Finally, a socioeconomic study of the conditions leading to prefer a strategy based on high-degree or on low-degree nodes would perhaps produce new insights with respect to practical constraints for adaptive network controllability.  

\bibliographystyle{abbrv}
\bibliography{ComplexNw2016}

\begin{thebibliography}{10}

\bibitem{adamopoulos2015unexpectedness}
P.~Adamopoulos and A.~Tuzhilin.
\newblock On unexpectedness in recommender systems: Or how to better expect the
  unexpected.
\newblock {\em ACM Transactions on Intelligent Systems and Technology (TIST)},
  5(4):54, 2015.

\bibitem{adomavicius2012improving}
G.~Adomavicius and Y.~Kwon.
\newblock Improving aggregate recommendation diversity using ranking-based
  techniques.
\newblock {\em IEEE Transactions on Knowledge and Data Engineering},
  24(5):896--911, 2012.

\bibitem{ACC-compleNet-2014}
L.~Allodi, L.~Chiodi, and M.~Cremonini.
\newblock Self-organizing techniques for knowledge diffusion in dynamic social
  networks.
\newblock In {\em Proceedings of {C}omplex {N}etworks {C}onference 2014
  ({C}omple{N}et14)}. Bologna, Italy, 2014.

\bibitem{barrat2004architecture}
A.~Barrat, M.~Barthelemy, R.~Pastor-Satorras, and A.~Vespignani.
\newblock The architecture of complex weighted networks.
\newblock {\em Proceedings of the National Academy of Sciences of the United
  States of America}, 101(11):3747--3752, 2004.

\bibitem{bozdag2013bias}
E.~Bozdag.
\newblock Bias in algorithmic filtering and personalization.
\newblock {\em Ethics and information technology}, 15(3):209--227, 2013.

\bibitem{bradley2001improving}
K.~Bradley and B.~Smyth.
\newblock Improving recommendation diversity.
\newblock In {\em Proceedings of the Twelfth Irish Conference on Artificial
  Intelligence and Cognitive Science, Maynooth, Ireland}, pages 85--94.
  Citeseer, 2001.

\bibitem{casamassima2016use}
F.~Casamassima and M.~Cremonini.
\newblock Use of random topics as practical control signals in a social network
  model.
\newblock In {\em International Workshop on Complex Networks and their
  Applications}, pages 539--550. Springer, 2016.

\bibitem{centola2007homophily}
D.~Centola, J.~C. Gonzalez-Avella, V.~M. Eguiluz, and M.~San~Miguel.
\newblock Homophily, cultural drift, and the co-evolution of cultural groups.
\newblock {\em Journal of Conflict Resolution}, 51(6):905--929, 2007.

\bibitem{cc-opres-2015}
A.~Ceselli and M.~Cremonini.
\newblock Models and methods for the analysis of the diffusion of skills in
  social networks.
\newblock In {\em Proceedings of {I}nternational {C}onference in {O}peration
  {R}esearch ({OR}2015)}. Wien, Austria, 2015.

\bibitem{chen2015paradox}
Y.-Z. Chen, L.~Wang, W.~Wang, and Y.-C. Lai.
\newblock The paradox of controlling complex networks: control inputs versus
  energy requirement.
\newblock {\em arXiv preprint arXiv:1509.03196}, 2015.

\bibitem{cornelius2013realistic}
S.~P. Cornelius, W.~L. Kath, and A.~E. Motter.
\newblock Realistic control of network dynamics.
\newblock {\em Nature communications}, 4, 2013.

\bibitem{cowan2012nodal}
N.~J. Cowan, E.~J. Chastain, D.~A. Vilhena, J.~S. Freudenberg, and C.~T.
  Bergstrom.
\newblock Nodal dynamics, not degree distributions, determine the structural
  controllability of complex networks.
\newblock {\em PloS one}, 7(6):e38398, 2012.

\bibitem{crandall2008feedback}
D.~Crandall, D.~Cosley, D.~Huttenlocher, J.~Kleinberg, and S.~Suri.
\newblock Feedback effects between similarity and social influence in online
  communities.
\newblock In {\em Proceedings of the 14th ACM SIGKDD international conference
  on Knowledge discovery and data mining}, pages 160--168. ACM, 2008.

\bibitem{cremonini2016introducing}
M.~Cremonini.
\newblock Introducing serendipity in a social network model of knowledge
  diffusion.
\newblock {\em Chaos, Solitons \& Fractals}, 90:64--71, 2016.

\bibitem{da2015sudden}
J.~da~Gama~Batista, J.-P. Bouchaud, and D.~Challet.
\newblock Sudden trust collapse in networked societies.
\newblock {\em The European Physical Journal B}, 88(3):55, 2015.

\bibitem{delellis2017evolving}
P.~DeLellis, A.~DiMeglio, F.~Garofalo, and F.~Lo~Iudice.
\newblock The evolving cobweb of relations among partially rational investors.
\newblock {\em PloS one}, 12(2):e0171891, 2017.

\bibitem{easley2012networks}
D.~Easley, J.~Kleinberg, et~al.
\newblock Networks, crowds, and markets: Reasoning about a highly connected
  world.
\newblock {\em Significance}, 9:43--44, 2012.

\bibitem{ehrhardt2006phenomenological}
G.~C. Ehrhardt, M.~Marsili, and F.~Vega-Redondo.
\newblock Phenomenological models of socioeconomic network dynamics.
\newblock {\em Physical Review E}, 74(3):036106, 2006.

\bibitem{gao2014target}
J.~Gao, Y.-Y. Liu, R.~M. D'Souza, and A.-L. Barab{\'a}si.
\newblock Target control of complex networks.
\newblock {\em Nature communications}, 5, 2014.

\bibitem{Goldstone2009}
R.~L. Goldstone and T.~M. Gureckis.
\newblock Collective behavior.
\newblock {\em Topics in Cognitive Science}, 1(3):412--438, 2009.

\bibitem{golub2012homophily}
B.~Golub and M.~O. Jackson.
\newblock How homophily affects the speed of learning and best response
  dynamics.
\newblock 2012.

\bibitem{grigoriev1997pinning}
R.~Grigoriev, M.~Cross, and H.~Schuster.
\newblock Pinning control of spatiotemporal chaos.
\newblock {\em Physical Review Letters}, 79(15):2795, 1997.

\bibitem{gross2008adaptive}
T.~Gross and B.~Blasius.
\newblock Adaptive coevolutionary networks: a review.
\newblock {\em Journal of the Royal Society Interface}, 5(20):259--271, 2008.

\bibitem{gross2006epidemic}
T.~Gross, C.~J.~D. D'Lima, and B.~Blasius.
\newblock Epidemic dynamics on an adaptive network.
\newblock {\em Physical review letters}, 96(20):208701, 2006.

\bibitem{iaquinta2008introducing}
L.~Iaquinta, M.~De~Gemmis, P.~Lops, G.~Semeraro, M.~Filannino, and P.~Molino.
\newblock Introducing serendipity in a content-based recommender system.
\newblock In {\em Hybrid Intelligent Systems, 2008. HIS'08. Eighth
  International Conference on}, pages 168--173. IEEE, 2008.

\bibitem{iniguez2009opinion}
G.~Iniguez, J.~Kert{\'e}sz, K.~K. Kaski, and R.~A. Barrio.
\newblock Opinion and community formation in coevolving networks.
\newblock {\em Physical Review E}, 80(6):066119, 2009.

\bibitem{kleinsman2015facebook}
J.~Kleinsman and S.~Buckley.
\newblock Facebook study: a little bit unethical but worth it?
\newblock {\em Journal of Bioethical inquiry}, 12(2):179--182, 2015.

\bibitem{kossinets2006empirical}
G.~Kossinets and D.~J. Watts.
\newblock Empirical analysis of an evolving social network.
\newblock {\em science}, 311(5757):88--90, 2006.

\bibitem{li2014sparse}
M.~Li, S.~Guan, C.~Wu, X.~Gong, K.~Li, J.~Wu, Z.~Di, and C.-H. Lai.
\newblock From sparse to dense and from assortative to disassortative in online
  social networks.
\newblock {\em Scientific reports}, 4, 2014.

\bibitem{lin1974structural}
C.-T. Lin.
\newblock Structural controllability.
\newblock {\em IEEE Transactions on Automatic Control}, 19(3):201--208, 1974.

\bibitem{liu2015control}
Y.-Y. Liu and A.-L. Barab{\'a}si.
\newblock Control principles of complex networks.
\newblock {\em arXiv preprint arXiv:1508.05384}, 2015.

\bibitem{liu2011controllability}
Y.-Y. Liu, J.-J. Slotine, and A.-L. Barab{\'a}si.
\newblock Controllability of complex networks.
\newblock {\em Nature}, 473(7346):167--173, 2011.

\bibitem{iudice2015structural}
F.~Lo~Iudice, F.~Garofalo, and F.~Sorrentino.
\newblock Structural permeability of complex networks to control signals.
\newblock {\em Nature communications}, 6, 2015.

\bibitem{lu2016attack}
Z.-M. Lu and X.-F. Li.
\newblock Attack vulnerability of network controllability.
\newblock {\em PloS one}, 11(9):e0162289, 2016.

\bibitem{maglajlic2012engineering}
S.~Maglajlic.
\newblock Engineering social networks using the controllability approach
  applied to e-learning.
\newblock In {\em Advanced Learning Technologies (ICALT), 2012 IEEE 12th
  International Conference on}, pages 276--280. IEEE, 2012.

\bibitem{marceau2010adaptive}
V.~Marceau, P.-A. No{\"e}l, L.~H{\'e}bert-Dufresne, A.~Allard, and L.~J.
  Dub{\'e}.
\newblock Adaptive networks: Coevolution of disease and topology.
\newblock {\em Physical Review E}, 82(3):036116, 2010.

\bibitem{mcpherson2001birds}
M.~McPherson, L.~Smith-Lovin, and J.~M. Cook.
\newblock Birds of a feather: Homophily in social networks.
\newblock {\em Annual review of sociology}, pages 415--444, 2001.

\bibitem{menichetti2014network}
G.~Menichetti, L.~Dall'Asta, and G.~Bianconi.
\newblock Network controllability is determined by the density of low in-degree
  and out-degree nodes.
\newblock {\em Physical review letters}, 113(7):078701, 2014.

\bibitem{motter2015networkcontrology}
A.~E. Motter.
\newblock Networkcontrology.
\newblock {\em Chaos: An Interdisciplinary Journal of Nonlinear Science},
  25(9):097621, 2015.

\bibitem{newman2010networks}
M.~Newman.
\newblock {\em Networks: an introduction}.
\newblock Oxford university press, 2010.

\bibitem{newman2002assortative}
M.~E. Newman.
\newblock Assortative mixing in networks.
\newblock {\em Physical review letters}, 89(20):208701, 2002.

\bibitem{newman2003mixing}
M.~E. Newman.
\newblock Mixing patterns in networks.
\newblock {\em Physical Review E}, 67(2):026126, 2003.

\bibitem{newman:2003}
M.~E. Newman.
\newblock The structure and function of complex networks.
\newblock {\em SIAM Review}, 45(2):167--256, 2003.

\bibitem{newman2003social}
M.~E. Newman and J.~Park.
\newblock Why social networks are different from other types of networks.
\newblock {\em Physical Review E}, 68(3):036122, 2003.

\bibitem{pariser2011filter}
E.~Pariser.
\newblock {\em The filter bubble: What the Internet is hiding from you}.
\newblock Penguin UK, 2011.

\bibitem{pennacchioli2013three}
D.~Pennacchioli, G.~Rossetti, L.~Pappalardo, D.~Pedreschi, F.~Giannotti, and
  M.~Coscia.
\newblock The three dimensions of social prominence.
\newblock In {\em International Conference on Social Informatics}, pages
  319--332. Springer, 2013.

\bibitem{sorrentino2007controllability}
F.~Sorrentino, M.~di~Bernardo, F.~Garofalo, and G.~Chen.
\newblock Controllability of complex networks via pinning.
\newblock {\em Physical Review E}, 75(4):046103, 2007.

\bibitem{sun2015controllability}
P.~G. Sun.
\newblock Controllability and modularity of complex networks.
\newblock {\em Information Sciences}, 325:20--32, 2015.

\bibitem{tucker2014social}
C.~E. Tucker.
\newblock Social networks, personalized advertising, and privacy controls.
\newblock {\em Journal of Marketing Research}, 51(5):546--562, 2014.

\bibitem{vardaman2012interpreting}
J.~M. Vardaman, J.~M. Amis, B.~P. Dyson, P.~M. Wright, and R.~Van~de
  Graaff~Randolph.
\newblock Interpreting change as controllable: The role of network centrality
  and self-efficacy.
\newblock {\em Human Relations}, 65(7):835--859, 2012.

\bibitem{wang2002pinning}
X.~F. Wang and G.~Chen.
\newblock Pinning control of scale-free dynamical networks.
\newblock {\em Physica A: Statistical Mechanics and its Applications},
  310(3):521--531, 2002.

\bibitem{white2013beliefs}
R.~White.
\newblock Beliefs and biases in web search.
\newblock In {\em Proceedings of the 36th international ACM SIGIR conference on
  Research and development in information retrieval}, pages 3--12. ACM, 2013.

\bibitem{yao2017structural}
P.~Yao, B.-Y. Hou, Y.-J. Pan, and X.~Li.
\newblock Structural controllability of temporal networks with a single
  switching controller.
\newblock {\em PloS one}, 12(1):e0170584, 2017.

\bibitem{zhang2007expertise}
J.~Zhang, M.~S. Ackerman, and L.~Adamic.
\newblock Expertise networks in online communities: structure and algorithms.
\newblock In {\em Proceedings of the 16th international conference on World
  Wide Web}, pages 221--230. ACM, 2007.

\end{thebibliography}

% % Non-BibTeX users please use
% \begin{thebibliography}{}
% %
% % and use \bibitem to create references. Consult the Instructions
% % for authors for reference list style.
% %
% \bibitem{RefJ}
% % Format for Journal Reference
% Author, Article title, Journal, Volume, page numbers (year)
% % Format for books
% \bibitem{RefB}
% Author, Book title, page numbers. Publisher, place (year)
% % etc
% \end{thebibliography}

\end{document}